\documentclass[twoside]{IEEEtran}
\usepackage{graphicx}
\usepackage{amsmath}
\usepackage{mathptmx} 
\usepackage{cite} 
\begin{document} 
\twocolumn[
\thispagestyle{empty}
\noindent
© 2020 IEEE. Personal use of this material is permitted. Permission from IEEE must be obtained
for all other uses, in any current or future media, including reprinting/republishing this material for
advertising or promotional purposes, creating new collective works, for resale or redistribution to
servers or lists, or reuse of any copyrighted component of this work in other works.
]
\newpage
\setcounter{page}{0}
\title{Contribution of residual quasiparticles to the characteristics of superconducting thin-film resonators
\author{T. Noguchi, S. Mima, and C. Otani 
\thanks{Manuscript received September 30, 2018. }
\thanks{T. Noguchi is with the Advanced Technology Center, National Astronomical Observatory, 2-1-1 Osawa, Mitaka, Tokyo, 181-8588 Japan (email: Takashi.Noguchi@nao.ac.jp).} 
\thanks{T. Noguchi, S. Mima, and C. Otani are  with RIKEN, , 2–1 Hirosawa, Wako, Saitama 351–0198, Japan. }  
}} 
\maketitle
\markboth
{}{Noguchi \MakeLowercase{\textit{et al.}}: Contribution of residual quasiparticles to the characteristics of superconducting thin-film resonators} 
\begin{abstract} 
We found that there are a significant number of quasiparticles present in the superconductor even at $T \ll T_c$ and that those quasiparticles seriously contribute to the characteristics of superconducting thin-film resonators.
The temperature behaviors of the resonator characteristics are well explained by that of the normal conductivity of quasiparticles including the contributions from Kondo effect, phonon, and electron-electron scatterings.
The readout microwave power dependence of the resonator characteristics are well explained by the change of quasiparticle density due to the energy re-distribution of quasiparticles under microwave field in the resonator. 
The resistive loss of the resonator decreases as the power of the readout microwave increases, and when the power of the readout microwave becomes sufficiently large, the resonator loss is dominated by the dielectric loss of the substrate.
\end{abstract}


\maketitle 
\IEEEdisplaynontitleabstractindextext 
\IEEEpeerreviewmaketitle 
\section{Introduction} 
\IEEEPARstart  
We have been developing Microwave Kinetic Inductance detector (MKID) for future application to large format radio cameras for astronomical observations.
Now we are making and studying two kinds of MKIDs; one is based on Al quarter-wave resonator and the other is Nb quarter-wave resonator \cite{Day, Mazin}.
and their quality factors and
resonance frequencies with respect to not only temperature but also readout
microwave power are being measured. 

\begin{figure}[t]
\begin{center}
\includegraphics[width=0.8\linewidth]{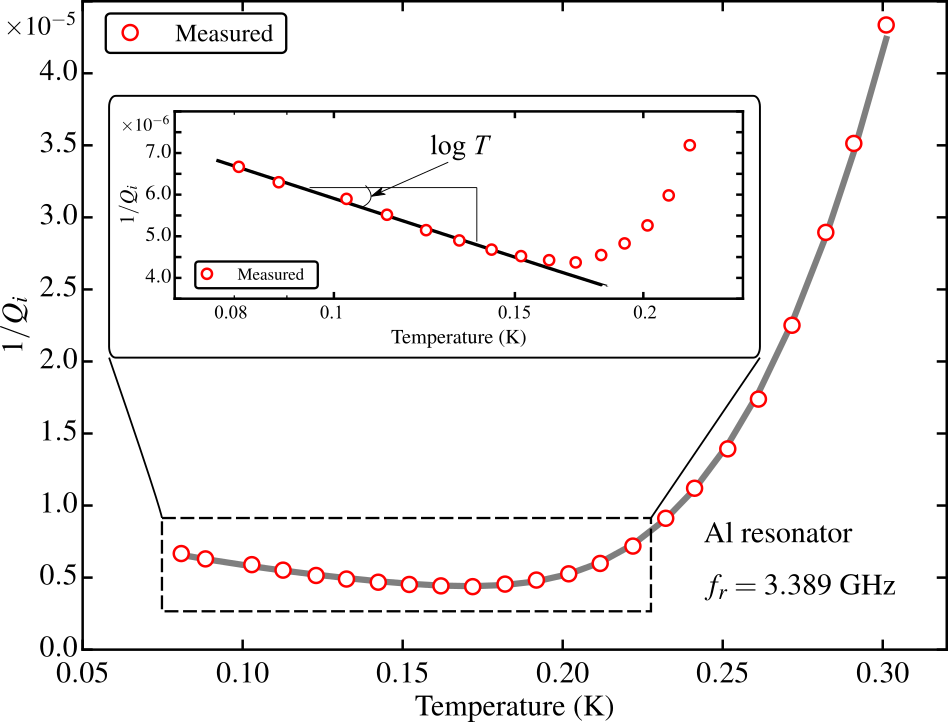}
\end{center}
\caption{\label{fig1} (a) Measured Internal quality factor $Q_i$ (upper panel) and fractional resonance frequency change $\delta f_r / f_r$  (lower panel) as a function of temperature. Red lines are the calculated ones  to fit the measured data. (b) $\log T$ dependence of fractional resonance frequency change $\delta f_r / f_r$.}
\end{figure}

When we measured the internal quality factor of a resonator at temperatures well below $T_c$, we frequently found that $Q_i$ decreases (or $1/Q_i$ increases) as temperature is lowered as shown in Fig.~\ref{fig1}. It is interesting that $1/Q_i$ increases as $\log T$, which is shown in the inset of Fig.~\ref{fig1}.
Since the $1/Q_i$ is approximately equal to the ratio of residual resistance to total reactance of the resonator, $\log T$ dependence of $1/Q_i$ indicates that residual resistance is proportional to $\log T$, which strongly reminds me of the Kondo effect in the normal metal.
The temperature behavior of $1/Q_i$ indicates that there are significant number of quasiparticles present even at $T \ll T_c$ and that they behave  like electrons in normal metals.

So, at the beginning of this research we made the following assumptions:
\begin{enumerate}
\item 
There are significant number of quasiparticles in the superconductor even at temperatures well below $T_c$.
\item 
Those residual quasiparticles in the superconductor behave like electrons in a normal metal under a microwave field.
\item 
The residual resistance of quasiparticles in the superconductor depends not only on temperature but also on readout microwave power even at the frequency much smaller than the gap frequency.
\end{enumerate}
Then, we measured and analyzed the temperature dependence of the internal quality factor and the resonance frequency using a fitting method  based on both the extended Mattis-Bardeen theory \cite{noguchi2016, M-B} and the residual resistance taking into account the contribution of scatterings of the quasiparticles at low temperature. 
In this paper, it will be shown that the temperature behavior of not
only  the internal quality factor but also the resonance frequency
observed in the Al and Nb film resonators are determined by the
residual resistance contributed from quasiparticle scatterings such as
the Kondo effect at low temperature.

It will be also shown that the energy redistribution of the
 residual quasiparticles is induced due to a strong microwave field in the
 resonator, so that the number of quasiparticles near the Fermi level,
 which mainly contribute to the normal conductivity, decreases with
 increasing the microwave field. It will be shown that
 the increase of the internal quality factor of the resonator with
 increasing readout microwave power might be attributed to the energy
 redistribution of the quasiparticles under the strong microwave field
 in the resonator. Finally, it will be mentioned that the loss of the resonator at the high readout microwave power is dominated by the contribution of the dielectric loss of the substrate.
 
\begin{figure}[t]
\begin{center}
\includegraphics[height=5cm]{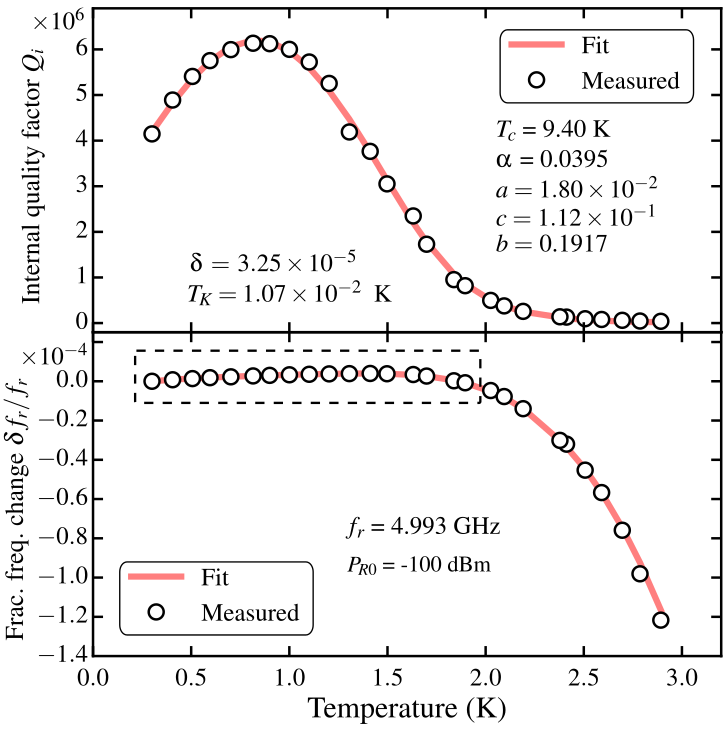}
\hspace{0.01\linewidth}
\includegraphics[height=5cm]{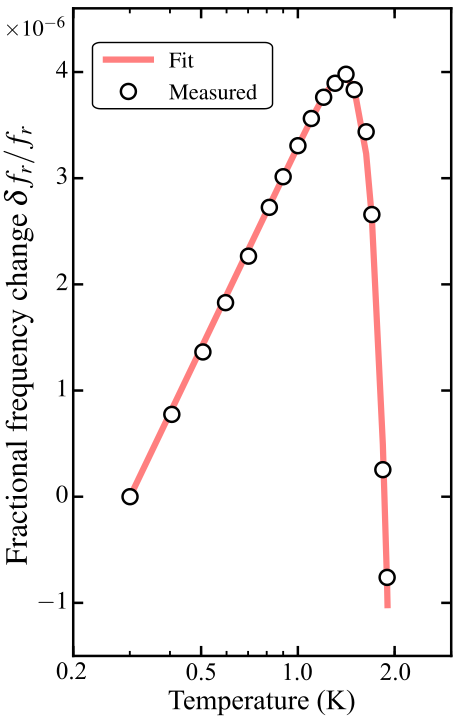} \\
\hspace*{0.15\linewidth}(a) \hspace{0.45\linewidth}(b) 
\end{center}
\caption{\label{fig2} (a) Measured Internal quality factor $Q_i$ (upper panel) and fractional resonance frequency change $\delta f_r / f_r$  (lower panel) as a function of temperature. Red lines are the calculated ones  to fit the measured data. (b) $\log T$ dependence of fractional resonance frequency change $\delta f_r / f_r$.}
\end{figure}

\section{Experimental}
Quarter wavelength coplanar waveguide (CPW) type thin-film resonators
using Al and Nb film were made onto a 4$\times$20 mm$^2$ Si and
sapphire chips respectively.  The width of a center line of the CPW and the gap
between the center line and the ground conductor were 8 and 3 $\mu$m,
respectively.
A microwave signal in the 3.8 or 4.9 GHz band is transmitted from one
port of the CPW feed line (readout line) to the other port and the
transmission coefficient $S_{21}$ was measured as a function of
temperature and frequency using a vector network analyzer (VNA).

It has been shown \cite{noguchi2018, noguchi2019} that
both the decrease of the $Q_i$ and the broad peak of the $\delta f_r /
f_r$ at low temperature in the Nb resonator can be well explained by the
increase of residual resistance of quasiparticles due to the Kondo
effect \cite{Kondo_effect}.  Light red lines in upper and lower panel of the left figure in Fig.~\ref{fig2} are fitting curves calculated by the eqs.~(4) and (5) in
Ref.\cite{noguchi2019} taking into account the contributions of the
Kondo effect, phonon scattering and electron-electron scattering to the
residual resistance of the quasiparticles. It should be also noted here
that the broad peak of $\delta f_r / f_r$ around 1.5 K is caused by the
contribution of the kinetic inductance of the residual quasiparticles.
%
\begin{figure}[t]
\begin{center}
\includegraphics[width=0.8\linewidth]{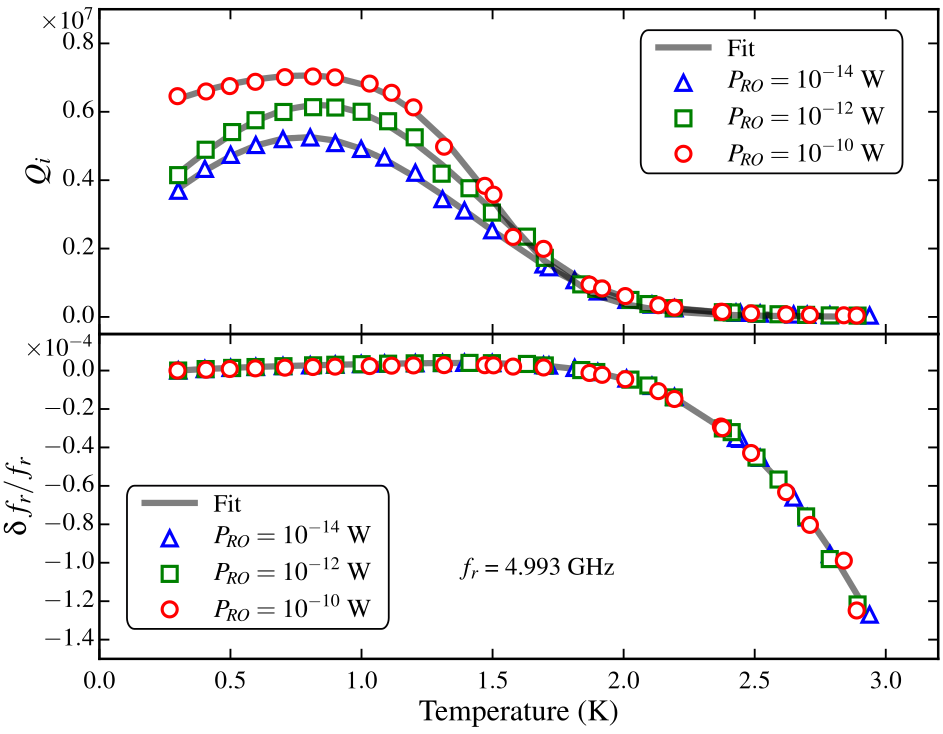}
\end{center}
\caption{\label{fig3} Internal quality factor $Q_i$ (upper panel) and
 resonance frequency $f_r$ (lower panel) of the Nb resonator as a function of
temperature for different readout microwave power $P_{RO}$.}
\end{figure}

From the fact that both the measured $1/Q_i$ and $\delta f_r/f_r$ agree
very well with those calculated by taking into account the contribution
of the residual resistance of quasiparticles at low temperature, we
confirmed that a significant number of quasiparticles still remain in
the superconducting gap at low temperature and that they behaves like
the electrons in normal metals.
Not only the temperature but also the readout microwave power dependences of $Q_i$ and $\delta f_r/f_r$ of a Nb resonator are shown in Fig.~\ref{fig3}.
It is shown that our approximate expressions for $Q_i$ and $\delta f_r/f_r$ are well fit to the measured ones for different readout microwave power.
Interesting point of this plot is that $Q_i$ strongly depends on readout microwave power, whereas $\delta f_r/f_r$ is almost independent of the readout microwave power. 
Very similar behavior of $Q_i$ and $f_r$ with respect to the readout
microwave power $P_{RO}$ was also observed in the Al resonator as shown
in the upper and lower panel in Fig.~\ref{fig4}, respectively.  
It should be noted here that
the $Q_i$ of the Al resonator changes more than factor of 10 in
magnitude in the similar power range of $P_{RO}$ at the lowest measured
temperature., whereas $f_r$ does not depend on the readout microwave power $P_{RO}$. 
In the next section, we will discuss the reason why $Q_i$ increases
with increasing readout microwave power while the resonance frequency
remains constant.

\section{Readout microwave power dependence}
\begin{figure} [t]
\begin{center}
\includegraphics[width=0.8\linewidth]{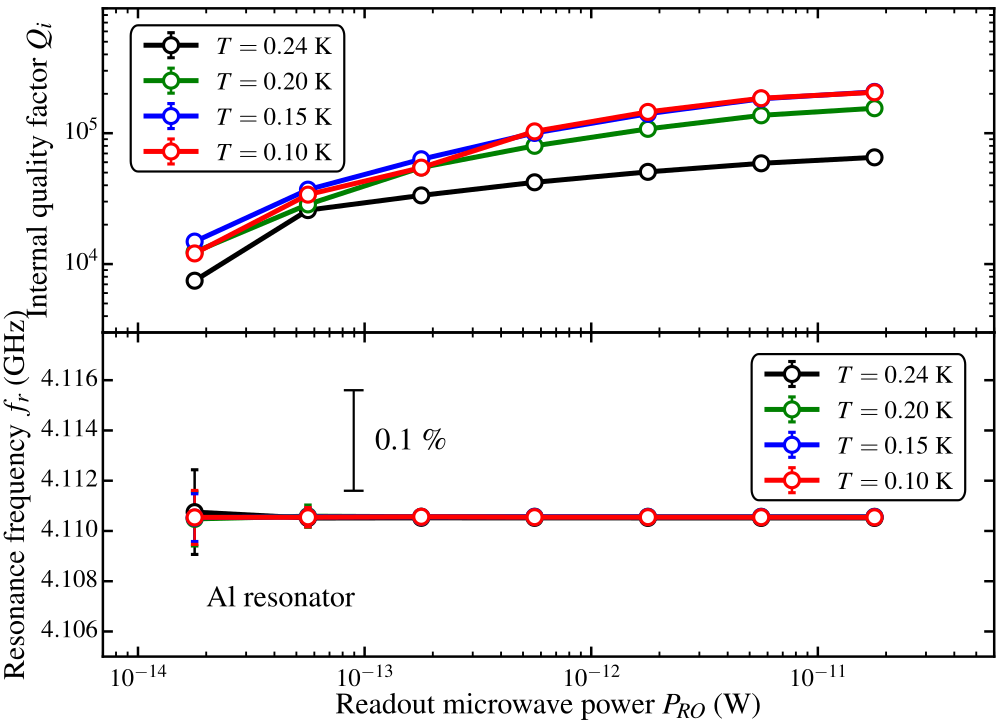}
\end{center}
\caption{\label{fig4}  Internal quality factor $Q_i$ (upper panel) and
resonance frequency $f_r$ (lower panel) of the Al resonator as a function of readout microwave
power $P_{RO}$ for different temperatures. }
\end{figure}

It is clearly shown by Fig.~\ref{fig4}
that the resonance frequency $f_r$ for the Nb and Al resonator are
almost constant and independent of the readout microwave power
$P_{RO}$. Assuming that the microwave properties of a superconductor can be described by a complex conductivity, $\sigma = \sigma_1 - i\sigma_2$, this means that the $\sigma_2$ of the superconducting Al and
Nb are approximately unchanged against the $P_{RO}$ 
because the resonance frequency $f_r$ is determined only by  $\sigma_2$.
On the other hand, the internal quality factor $Q_i$ of the Al resonator
are strongly dependent on the readout microwave power $P_{RO}$. It is found that 
the internal quality factor $Q_i$ of the Nb resonator also increases as
the readout microwave power $P_{RO}$ increases.
 This means that the $\sigma_1$ decreases as the readout
microwave power $P_{RO}$ increases in this power range.  Because the
internal quality factor $Q_i$ of a resonator is described by a ratio of
the $\sigma_2$ and $\sigma_1$ of the superconductor and the $\sigma_2$ is
constant against the readout microwave power $P_{RO}$ in this power range as mentioned above.
%

According to the Drude model \cite{Ascroft_Mermin}, the quasiparticle current ${\bf J}_{qp}$
in the microwave electric field ${\bf E}$ is a function of the
conductivity $\sigma_1$ as
\begin{equation}
{\bf J}_{qp} = \frac{n_{qp} e^2 \tau}{m} {\bf E} = \sigma_1 {\bf E}, 
\end{equation}
where ${\bf E}$ is an amplitude of the microwave field, $n_{qp}$ is the
number of quasiparticles, $\tau$ is a scattering time, and $m$ and $e$
are the electron mass and charge, respectively. 
The conductivity
$\sigma_1$ is proportional to the number of quasiparticle $n_{qp}$ as
\begin{equation}
\sigma_1 = \frac{n_{qp} e^2 \tau}{m}  \label{sigma_1}. 
\end{equation}
According to eq.~(\ref{sigma_1}), the the inverse quality factor $1/Q_i$
of a superconducting resonator is thought to be proportional to the
quasiparticle density $n_{qp}$, since the $1/Q_i$ is proportional to the
conductivity $\sigma_1$ as described above.  Therefore, we calculated
the number of quasiparticles under the microwave field in the resonator,
assuming that the number of quasiparticles $n_{qp}$ in the resonator changes according to the readout microwave power $P_{RO}$.

Following Tien and Gordon's argument \cite{Tien_Gordon}, we start
with a very simple case that a microwave electric field is excited
between the center conductor and the ground plane of the CPW.  In this case, the microwave 
electric field sets up a potential difference $V_{res} \cos \omega t$ between
them.

When no field is present the wave functions of the charge carriers of
energy $\mathcal{E}$ satisfy the unperturbed Hamiltonian $H_0$
\begin{equation}
\psi({\bf{r}}, t) = \psi_0({\bf{r}}) \exp\left(- i \frac{\mathcal{E} t}{\hbar}\right) 
\end{equation}
The perturbed Hamiltonian due to the oscillating electric field is given by 
\begin{equation}
H = H_0 + e V_{res} \cos \omega t
\end{equation}
This interaction Hamiltonian only effects the time-dependent part of the
wave function. 

The new wave function under influence of the microwave field becomes
\begin{align}
\psi({\bf{r}}) &= \psi_0({\bf{r}}) \exp\left\{- i \frac{\mathcal{E}
t}{\hbar}-\frac{i}{\hbar} \int_0^t e V_{res} \cos (\omega
t') d t' \right\} \nonumber \\
  &= \psi_0({\bf{r}}) \exp\left(- i \frac{\mathcal{E} t}{\hbar}\right)
  \sum_{-\infty}^{\infty} J_n (\xi) \exp\left(- i n\omega t\right) ,
\end{align}
where $J_n(\xi)$ is the $n$-th order Bessel function of the first kind and
$\xi=\frac{e V_{res}}{\hbar\omega}$ is an amplitude of the normalized microwave
voltage.  The wavefunction of a quasiparticle is modulated by the perturbation of $eVrf cos \omega t$ and a quasiparticle state split into many levels with a energy spacing $\hbar\omega$ and an amplitude Bessel function $J_n(\xi)$.  Accordingly, the effective
density of states in the presence of the microwave field becomes
\begin{equation}
D^{\prime} (\mathcal{E}) = \sum_{n=-\infty}^{\infty}D_s(\mathcal{E} + n\hbar\omega)   J_n^2(\xi) ,
\end{equation}
where $D_s(\mathcal{E}) $ is an unperturbed density of states in a
superconductor given by
\begin{equation}
D_s(\mathcal{E})  = \left| \frac{\mathcal{E}}{\sqrt{\mathcal{E}^2 - \Delta^2}}\right| .
\end{equation} 
where $\Delta$ is a complex gap energy of the superconductor, {\it i.e.} $\Delta =\Delta_1 + i\Delta_2$ and $\frac{\Delta_2}{\Delta_1} \ll 1$ and $\Delta_1$,  and $\Delta_2$ are real numbers, respectively \cite{noguchi2016}. 
%
\begin{figure}[t]
\begin{center}
\includegraphics[width=0.8\linewidth]{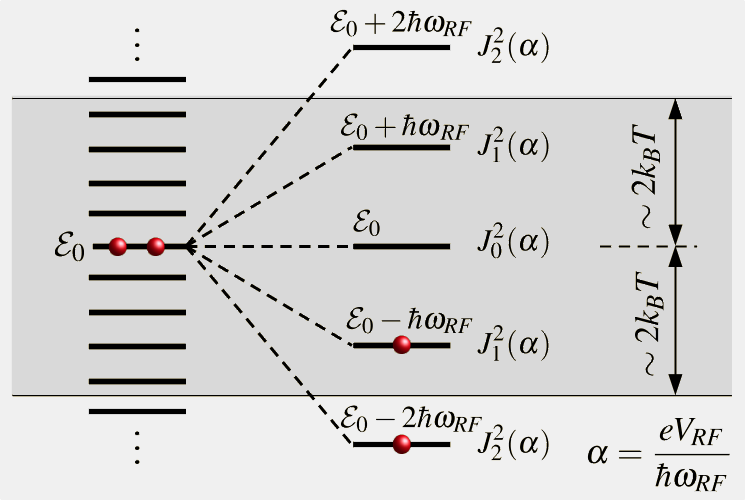}
\end{center}
\caption{\label{fig5} Development of sidebands of the original energy as a consequence of the microwave field. The hatched region corresponds to the tail of the Fermi distribution due to the thermal broadening.}
\end{figure}

When  split multi-levels are generated, some quasiparticles transition to lower energy levels, resulting in a change in the energy distribution of the quasiparticles and in a decrease of the number of quasiparticles occupying the tail of the Fermi function.
The quasiparticle density under microwave field is given by
\begin{align}
n_{qp}(x) &= 4 N_0\int_{0}^{\infty}D_s(\mathcal{E})
 f(\mathcal{E})\, d\mathcal{E} \nonumber \\
  &= n_{qp}(0)\, J_0^2(x) + 4
 N_0\sum_{n=1}^{\infty} J_n^2(\xi) \nonumber \\ 
  & \ \times \int_{0}^{\infty} [ D_{s}
 (\mathcal{E}-n\hbar\omega) +D_{s} (\mathcal{E} + n\hbar\omega) ]
 f(\mathcal{E}) \, d\mathcal{E} \label{n_qp1}
\end{align}
where $n_{qp}(0)$ is a quasiparticle density in the thermal equilibrium is defined by
\begin{equation}
n_{qp}(0) = 4 N_0\int_{0}^{\infty}D_{s0}(\mathcal{E}) f(\mathcal{E})\,  d\mathcal{E} ,
\end{equation}
where $f(\mathcal{E})$ is the Fermi function. 
When the photon energy of the readout microwave is much smaller than the gap
energy ($\frac{\hbar\omega}{\Delta} \ll 1$) , which is satisfied in most MKID resonators, the following approximation can be applied for small $n$
\begin{equation}
D_{s}(\mathcal{E} \pm n \hbar \omega) \approx 
D_{s}(\mathcal{E}) \pm (n \hbar \omega) \frac{d D_s(\mathcal{E})}{d \mathcal{E}}
\end{equation}
%
Energy redistribution of quasiparticles is caused by transitions between quasiparticle levels involving the absorption and emission of multiple photons. Since the probability of such multiple-photon transitions decreases rapidly as the number of photons increases, the terms with $n\le 4$  in eq.~(\ref{n_qp1}) may be omitted. The 
\begin{equation}
n_{qp}(\xi) \simeq n_{qp}(0) \left [ J_{0}^2 (\xi) + 2\sum_{n=1}^{n=3} J_{n}^2 (\xi) \right]  . \label{n_qp}
\end{equation}
Considering that the inverse quality factor $1/Q_i$ is proportional to
the quasiparticle density $n_{qp}(\xi)$, eq.~(\ref{n_qp}) shows that the
inverse quality factor $1/Q_i$ varies as a function of the sum of  $J_n^2 (\xi)$ with
respect to the normalized microwave voltage $\xi$. This indicates that
quasiparticle energy redistribution occurs in the presence of the
microwave field, resulting in a decrease in the number of quasiparticles
with energy within $ \pm k_B T $ from the Fermi level.
%
\begin{figure}
\centering
\includegraphics[width=0.8\linewidth]{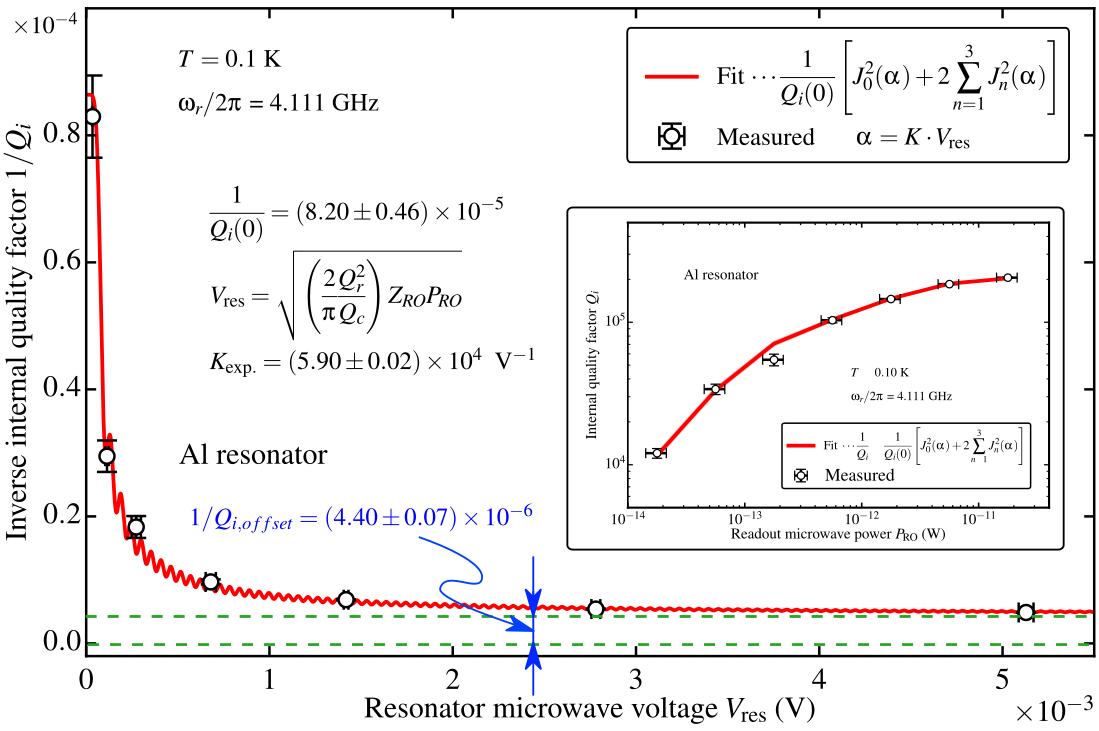} 
\caption{ \label{fig6} The inverse internal quality factors of $1/Q_i$ of the Al resonator at $T=0.1$  K are shown as a function of microwave voltage in the transmission-line resonator, $V_{res}$. The solid red line is calculate by eq.~(\ref{n_qp}). Inset is a plot of $Q_i$ as a function of the readout microwave power, $P_{RO}$. }
\end{figure}

Since the internal microwave power, $P_{INT}$, in a resonator is related to the readout microwave power $P_{RO}$ by \cite{Czakon}
\begin{equation}
P_{INT} = \left(\frac{2}{\pi}\frac{Q^2_r}{Q_c}\right)P_{RO},
\end{equation}
an amplitude of the microwave voltage
$V_{res}$ in the resonator transmission line is calculated as
\begin{equation}
V_{res} = \sqrt{\left(\frac{2}{\pi}\frac{Q^2_r}{Q_c}\right) Z_{RO} P_{RO}} , \label{V_res}
\end{equation}
where $Q_c$ and $Q_r$ are the coupling and loaded quality factor of the  resonator, respectively, and $Z_{RO}$ and $Z_{res}$ are 
characteristic impedance of a readout transmission line and the resonator transmission line, respectively and are assumed to be $Z_{res} = Z_{RO}\simeq 50\ \Omega$.

In Fig.~\ref{fig6}, measured inverse internal quality factor 
$1/Q_i$ of an Al resonator at  0.1 K is shown as a
function of the microwave voltage in the resonator calculate from the
measured readout microwave power $P_{RO}$ using
eq.~(\ref{V_res}). 

Note here that measured $1/Q_i$ seems, however, to have an offset of $1/Q_{i, offset}\approx 4.40\times 10^{-6}$ 
which is approximately only 5 \% of $1/Q_i$ for the lowest readout microwave power, and seems to be insensitive and constant against the microwave voltage
in the resonator.  Then we subtracted the constant offset from the measured $1/Q_i$ and fit to the
subtracted data was made using eq.~(\ref{n_qp}). The normalized voltage $\xi$ in the resonator is experimentally determined from the relation, $\xi = K\cdot V_{res}$, 
where $K$ is a scaling factor which is theoretically given by
$\displaystyle{\frac{e}{\hbar\omega_r}=5.87\times 10^4\ }$V$^{-1}$, where $\omega_r$ is
a resonance frequency of the resonator. It is clearly demonstrated in Fig~\ref{fig6} that
the measured data agree well to the calculated one using eq.~(\ref{n_qp}).  From the
fit we obtained the scaling parameter of $K=(5.90\pm0.02)\times 10^4\ V^{-1}$
which is only a factor 2 different from the theoretical value, which is quite consistent with the theoretical value of $\displaystyle{\frac{e}{\hbar\omega_r}=5.87\times 10^4}\ $V$^{-1}$.  
A very similar result of inverse quality factor $1/Q_i$ as a function of the normalized microwave voltage in the resonator is obtained in the Nb resonator measured at 0.8 K.
These fit results indicate that the quasiparticle energy
redistribution is caused by a strong microwave field in the resonator,
so that the number of quasiparticles near the Fermi level, which mainly
contribute to the normal conductivity $\sigma_1$, decreases with
increasing the amplitude of the microwave field.
%
\begin{table}  
\caption{Measured $1/Q_{i,offset}$ and $\tan \delta$ for Si and sapphire substrate }
\centering
\begin{tabular}{l|c|c} \hline 
Substrate material & $1/Q_{i,offset}$ (Measured) & $\tan \delta$ (Reported) \\ 
\hline \hline 
High resistivity Si &\parbox{2cm}{$1.09\times 10^{-6}$ \\ $4.40\times 10^{-6}$}  & $0.3 - 5 \times 10^{-6}$ \cite{si}\\ [1.5ex] \hline 
Crystalline sapphire & \parbox{2cm}{$2.43\times 10^{-8}$ \\ $1.38\times 10^{-7}$}  &  $0.8 - 5 \times 10^{-7}$ \cite{sap}  \\ 
  \hline
\end{tabular}
\label{tbl1} 
\end{table}

\section{Dielectric loss of the substrate}
Since the resonator used in the experiments is made of a quarter-wavelength CPW transmission line, the internal quality factor of the resonator, $Q_i$ is approximately equal the  quality factor of the transmission line, $Q_{TL.}$. 
From the viewpoint of the superconducting transmission line, the inverse of the quality factor of the transmission line, $1/Q_{TL}$ is given by the sum of the inverse of conductor and dielectric quality factors as
\begin{equation}
\frac{1}{Q_{TL}} = \frac{R_s}{\omega_r L} + \frac{G_d}{\omega_r C_d}
\approx \frac{1}{Q_{SC}} + \tan \delta 
\end{equation}
where $R_s$ and $\omega_r L$ are the surface resistance and reactance of the superconductor, respectively, $Q_{SC}$ is a quality factor of the superconductor, and $G_d$ and $\omega_r C_d$ and $\tan\delta$ are the conductance, susceptance and loss tangent of the dielectric, respectively. 
When the readout microwave power $P_{RO}$ is sufficiently high, the inverse   quality factor of the superconductor, $1/Q_{SC}$ approaches 0 and inverse transmission-line $Q_{TL}$ is dominated by the inverse dielectric quality factor, $\displaystyle \frac{G_d}{\omega_r C_d}$ or $\tan\delta$, which should be in close agreement with the $1/Q_{i,offset}$ obtained in the experiment, 
\begin{equation}
\frac{1}{Q_{i,offset}} \approx \frac{G_d}{\omega_r C_d} \approx \tan \delta
\end{equation}

The values of $1/Q_{i,offset}$ obtained for two Nb resonators on a sapphire substrate and and two Al resonators on  a Si substrate, respectively, are shown in Table \ref{tbl1}. In Table \ref{tbl1}, $\tan\delta$ of the sapphire and Si substrate measured at extremely low temperatures and reported are also shown. It is found that the measured $1/Q_{i,offset}$ for Nb and Al resonators obtained in our experiments are very consistent with reported loss tangent ($\tan\delta$) for Si and sapphire substrates measured at extremely low temperatures.   
\section{Summary}
It is demonstrated that there are a significant number of quasiparticles present in the superconductor even at $T \ll T_c$ and that those quasiparticles seriously contribute to the characteristics of superconducting thin-film resonators.

It is found that temperature behaviors of the resonator characteristics are well explained by that of the normal conductivity of quasiparticles including the contributions from Kondo effect, phonon, and electron-electron scatterings.
It is found that readout microwave power dependence of the resonator characteristics are well explained by the change of quasiparticle density due to the energy re-distribution of quasiparticles under microwave field in the resonator. 

The conductive loss of the resonator decreases as the power of the readout microwave increases, and when the power of the readout microwave becomes sufficiently large, the resonator loss is dominated by the dielectric loss of the substrate.

\end{document}